\documentclass[12pt]{iopart}
\usepackage{iopams}
\usepackage{graphics}
\begin{document}

\title[Thermodynamics of Time-Dependent  Wormholes at Event Horizon]{Modified $f(R)$ Gravity and Thermodynamics of Time-Dependent  Wormholes at Event Horizon}

\author{H. Saiedi } \ \\
\address{Department of Physics, Florida Atlantic University, FL 33431, USA  }
\ead{\mailto {hsaiedi2014@fau.edu \ ; \  hrssaiedi@gmail.com}}
\begin{abstract}\noindent \\
In the context of modified $f(R)$ gravity theory, we study time-dependent wormhole spacetimes in the radiation background. In this framework, we attempt to generalize the thermodynamic properties of  time-dependent  wormholes in $f(R)$ gravity. Finally, at event horizon, the rate of change of total entropy has been discussed.  \\ \\

\end{abstract}

\section{Introduction}
Wormholes, short cuts between otherwise distant or unconnected regions of the universe, have become a popular research topic since the influential paper
of Morris and Thorne [1]. Early work was reviewed in the book of Visser [2]. The Morris-Thorne (MT) study was restricted to static, spherically symmetric space-times, and initially there were various attempts to generalize the definition of wormhole by inserting time-dependent factors into the metric. According to the Einstein field equations, the MT wormhole needs an exotic matter (matter violating the null and weak energy conditions (NEC and WEC)) which holds up the wormhole structure and keeps the wormhole throat open. Sushkov[3] and Lobo[4] independently, have shown that phantom energy could well be the class of exotic matter which is required to support traversable wormholes.

Recent astronomical data indicate that the universe could be dominated by a fluid which violates the null energy condition[5-7]. Although, an exotic matter is responsible for the early-time inflation and late-time acceleration, the modified theory of gravity ($f(R)$ gravity), which can explain the present acceleration without introducing an exotic matter, has received intense attention. The literature on $f(R)$ gravity is vast, and the earlier reviews on this theory are given in [8-12]. \\

Since the discovery of black hole thermodynamics[13-15], physicists have been speculating that there should be some relationship between thermodynamics and Einstein equations. Jacobson is the first one who was able to derive Einstein equations from the proportionality of entropy to the horizon area of the black hole together with the first law of thermodynamics [16]. In the cosmological setup, Cai and his collaborators made the major development by showing that the Einstein field equations evaluated at the apparent horizon can also be expressed in the form of the first law of thermodynamics in various theories of gravity. This connection between gravity and thermodynamics has also been extended in the braneworld cosmology [17, 18]. All these indicate that the thermal interpretation of gravity is to be generic, so we investigate this relation for a more general spacetimes. The idea that wormholes may show some characteristics and properties which are parallel to those already found in black holes, seems to be quite natural, including in particular "wormhole thermodynamics" [19]. Also, the authors [20] have discussed wormhole thermodynamics at apparent horizons in Einstein gravity. \\ \\

\section{Thermodynamics of Time-dependent  Wormholes in  $f(R)$ Gravity}
In this section, we generalize the thermodynamics of evolving wormholes within the modified $f(R)$ gravity. Hayward first introduced a formalism for defining thermal properties of black holes in terms of measurable quantities. This formalism also works for the dynamical black holes which consistently recover the results obtained by global considerations using the event horizon as in the static case. A fascinating and rather surprising feature emerges if one recognize that the static wormhole reveals thermodynamics properties analogous to the black holes if one considers the local quantities. It is important to note that the non-vanishing surface gravity at the wormhole throat characterized by a non-zero temperature for which one would expect that wormhole should emit some sort of thermal radiation. \\
Let us start with the evolving wormhole metric[21] \\
\begin{equation}
ds^2 = - e^{2\Phi(t,r)} dt^2 + a^2(t) \left[ \frac{dr^2}{1 - \frac{b(r)}{r}
} +r^2 d\Omega^2_2  \right] \ ,
\end{equation} \ \\
where $d\Omega^2_2 = d\theta^2 +  \sin^2 \theta d\phi^2$ is the line element of a two-dimensional unit sphere, $a(t)$ is the scale factor of the universe, $b(r)$ and $ \Phi(t,r)$ are the arbitrary shape and redshift functions of the evolving wormhole, respectively. We rewrite the spherically symmetric metric (1) in the following form(by considering $ \Phi(t,r)=0$). \\
\begin{equation}
ds^2 = h_{ab} dx^a dx^b + \tilde{r}^2 d\Omega^2_2  \ \ \ \ , \ \ \ \ (a,b = 0, 1)
\end{equation} \ \\
where $x^0 = t, x^1 = r$, and $   \tilde{r}=a(t)r$ represents the radius of the sphere while the two-dimensional metric $h_{ab}$ is written as \\
\begin{equation}
h_{ab} = diag  \left[-1 ,  a^2(t) \left( 1 - \frac{b(r)}{r}  \right)^{-1} \right] \ .
\end{equation} \ \\ \\
The surface gravity is defined as [22, 23] \\
\begin{equation}
\kappa = \frac{1}{2\sqrt{-h}} \ \partial_a \left( \sqrt{-h} \  h^{ab} \partial_b \tilde{r}   \right) \ ,
\end{equation} \ \\
where $h$ is the determinant of metric $h_{ab}$  (3). So, the surface gravity at the wormhole horizon $\tilde{r}_h$ can be written as \\
\begin{equation}
\kappa = - \frac{\tilde{r}_h}{2} \left( \dot{H} + 2 H^2 \right) + \frac{1}{4\tilde{r}_h^2} \left( ab(r) - \tilde{r}_hb'(r) \right) \ ,
\end{equation} \ \\
where $b' = \partial{b}/\partial{r}$ and the overdot denotes differentiation with respect to time. $H=\dot{a}/a$ is the Hubble parameter. The horizon temperature is defined as $T_h = \kappa /2\pi$. So \\
\begin{equation}
T_h = - \frac{\tilde{r}_h}{4\pi} \left( \dot{H} + 2 H^2 \right) + \frac{1}{8\pi\tilde{r}_h^2} \left( ab(r) - \tilde{r}_hb'(r) \right) \ .
\end{equation} \ \\
The area of the wormhole horizon is defined as
\begin{equation}
A = 4\pi \tilde{r}_h^2
\end{equation} \ \\
One can relate the entropy with the surface area of the horizon through $S_h = \frac{A}{4G}\mid_{\tilde{r}_h}$. In the $f(R)$ theory of gravity, the entropy has the following relation to the horizon area \\
\begin{equation}
S_h = \frac{AF}{4G}\mid_{\tilde{r}_h} =  \frac{\pi \tilde{r}_h^2 F}{G} = 8\pi^2 \tilde{r}_h^2 F \ \ \ \ \ \ \ \ \ \ \ (8\pi G = 1) \ ,
\end{equation} \ \\
where  $ F = df/dR \neq 0$. Therefore \\
\begin{equation}
dS_h = 8\pi^2 \tilde{r}_h^2 dF + 16\pi^2 \tilde{r}_hF d\tilde{r}_h \ .
\end{equation} \ \\
Using the above equation and the relation (6), we obtain  \\
\begin{equation}
T_hdS_h =  \left(8\pi^2 \tilde{r}_h^2 dF + 16\pi^2 \tilde{r}_hF d\tilde{r}_h \right) \left[   \frac{ab(r) - \tilde{r}_hb'(r) }{8\pi\tilde{r}_h^2}   -  \frac{\tilde{r}_h \left( \dot{H} + 2 H^2 \right) }{4\pi}   \right] \ .
\end{equation} \ \\ \\
Now we consider the Gibbs equation [24] \\
\begin{equation}
T_h dS_I =  dE_I + p dV \ ,
\end{equation} \ \\
where $S_I$ is the entropy of the matter bounded by the horizon and $E_I$ is the energy of the matter distribution. The volume $V$ is defined as $V=\frac{4}{3}\pi\tilde{r}^3$, and $p$ denotes the average pressure inside the horizon which is $p = (p_r + 2p_t)/3$. Where $p_r(t,r)$ and $p_t(t,r)$ are the radial pressure and tangential pressure, respectively. Here for thermodynamical equilibrium, the temperature of the matter inside the horizon is assumed to be the same as on the horizon i.e. $T_h$. \\
Now starting with \\
\begin{equation}
V=\frac{4}{3}\pi\tilde{r}_h^3 \ \ \ \ \ \ , \ \ \ \ \ \ E_I = \rho V = \frac{4}{3}\pi\rho\tilde{r}_h^3 \ ,
\end{equation} \ \\
and the continuity equation $\dot{\rho} + H (3\rho + p_r + 2p_t) = 0$, which can be achieved from the energy conservation equation $T^{\mu}_{\nu ; \mu}$, the Gibbs equation leads to \\
\begin{equation}
T_h dS_I = \frac{4\pi \tilde{r}^2_h}{3} (3\rho + p_r + 2p_t + \frac{\tilde{r}_h\rho^{'}}{a})(d\tilde{r}_h - H\tilde{r}_h dt) \ ,
\end{equation} \ \\
where $\rho(t,r)$ is the energy density and $\rho^{'} = \partial \rho / \partial r$. \\
So, by combining equations (10) and (13), one can easily reach the following equation for the variation of total entropy. \\
\begin{eqnarray}
T_h \dot{S}_{tot}= T_h (\dot{S}_{I} + \dot{S}_{h}) &=&   - \frac{4\pi H\tilde{r}^3_h}{3} (3\rho + p_r + 2p_t + \frac{\tilde{r}_h\rho^{'}}{a})  \nonumber \\ \nonumber  \\              &+&                            8\pi^2 \tilde{r}_h^2\dot{F} \left[   \frac{ab(r) - \tilde{r}_hb'(r)       }{8\pi\tilde{r}_h^2}   -  \frac{\tilde{r}_h \left( \dot{H} + 2 H^2 \right) }{4\pi}   \right]    \nonumber \\ \nonumber  \\          &+&           16\pi^2 \tilde{r}_hF  \left[   \frac{ab(r) - \tilde{r}_hb'(r) }{8\pi\tilde{r}_h^2}   -  \frac{\tilde{r}_h \left( \dot{H} + 2 H^2 \right) }{4\pi}   \right] \dot{\tilde{r}}_h    \nonumber \\ \nonumber \\         &+&          \frac{4\pi\tilde{r}^2_h}{3} (3\rho + p_r + 2p_t +   \frac{\tilde{r}_h\rho^{'}}{a})\dot{\tilde{r}}_h  \  .
\end{eqnarray}  \\ \\

\section{Thermodynamics at Event Horizon}
Now we shall find out the event horizon radius and then analyze the above equation for evolving  wormholes. By considering $b(r)=r_0$, event horizon radius $\tilde{r}_E$ can be found from the relation(i.e. $ds^2 = 0 = d\Omega^2_2$) \\
\begin{equation}
\dot{\tilde{r}}_E = \tilde{r}_E H  -  \sqrt{ 1 - \frac{ar_0}{\tilde{r}_E} } \ ,
\end{equation}
or \\
\begin{equation}
\int_0^{\frac{\tilde{r}_E}{a}}  \frac{dr}{ \sqrt{ 1 - \frac{r_0}{r} }} \ = \int_t^{\infty} \frac{dt}{a} \ .
\end{equation} \ \\
Therefore, the event horizon radius can be found from the above equation. In the $f(R)$ theory of gravity, the author and Nasr Esfahani [25,26] have shown that the energy density $(\rho(t,r))$, radial pressure $(p_r(t,r))$ and tangential pressure $(p_t(t,r))$ can be written as \\
\begin{eqnarray}
\rho &=& -\ddot{F} +3H^2F   \ , \\
p_r &=& - 2\dot{H}F + H\dot{F} -3H^2F - \frac{r_0 F}{a^2r^3} \ , \\
p_t &=& - 2\dot{H}F + H\dot{F} -3H^2F + \frac{r_0F}{2a^2r^3}  \ ,
\end{eqnarray} \ \\
Now by substituting relations (17), (18), (19) and (15) into the equation (14), the variation of total entropy can be written as \\
\begin{eqnarray}
T_h \dot{S}_{tot} &=& - \frac{4\pi\tilde{r}^2_E}{3} \left(3H\dot{F} - 6\dot{H}F - 3\ddot{F} \right) \sqrt{ 1 - \frac{ar_0}{\tilde{r}_E} }   \nonumber \\ \nonumber \\       &+&        8\pi^2\tilde{r}_E \left( 2\tilde{r}_EHF + \tilde{r}_E\dot{F} \right) \left(\frac{ar_0}{8\pi \tilde{r}^2_E}   -   \frac{(\dot{H} + 2H^2)\tilde{r}_E  }{4\pi}  \right)   \nonumber  \\  \nonumber  \\  &-&         \left(16\pi^2\tilde{r}_E F \sqrt{ 1 - \frac{ar_0}{\tilde{r}_E} } \  \right)  \left(\frac{ar_0}{8\pi \tilde{r}^2_E}   -   \frac{(\dot{H} + 2H^2)\tilde{r}_E  }{4\pi}  \right) \ .
\end{eqnarray} \\ \\
Thus, the expression in equation (20) for the rate of change of total entropy at event horizon shows that the validity of the generalized second law of thermodynamics depends on the right hand side (r.h.s.) of the equation. The generalized second law will be valid for event horizon if the r.h.s. of the above expression is non-negative. \\ \\

\section{Conclusion}
We have reviewed the time-dependent  wormholes in modified $f(R)$ gravity. To support the wormhole geometry, we consider a traceless fluid.  We have generalized the thermodynamics of these  wormholes. By applying the area law of thermodynamics and using the Gibbs equation, the variation of total entropy at event horizon has been discussed. If this variation of total entropy is non-negative, the generalized second law of thermodynamics will be valid. It will be interesting in further investigation to study the  thermodynamic behavior of field equations in $f(R)$ gravity for rotating spacetimes.

\end{document}